\documentstyle[12pt,graphicx]{article}
\title{ Quintom models with an equation of state crossing $-1$ }
\author{\small Wen Zhao, Yang Zhang \\
        \small Astrophysics Center \\
        \small University of Science and Technology of China \\
        \small Hefei, Anhui, China }
 \date{}

\begin{document}
\maketitle
\baselineskip=19truept
\def\vek{\vec{k}}

\renewcommand\theequation{\arabic{equation}}
\newcommand{\be}{\begin{eqnarray}}
\newcommand{\ee}{\end{eqnarray}}
\newcommand{\ba}{\begin{eqnarray}}
\newcommand{\ea}{\end{eqnarray}}
\def\disp{\displaystyle}
\renewcommand{\H}{{\cal H}}
\renewcommand{\L}{{\cal L}}
\renewcommand{\P}{\tilde{\phi}}
\renewcommand{\t}{\tilde{\theta}}

\sf

\small

\begin{center}
\Large  Abstract
\end{center}
\begin{quote}
 {\small
In this paper, we investigate a kind of special quintom model,
which is made of a quintessence field $\phi_1$ and a phantom field
$\phi_2$, and the potential function has the form of
$V(\phi_1^2-\phi_2^2)$.  This kind of quintom fields can be
separated into two kinds: the hessence model, which has the state
of $\phi_1^2>\phi_2^2$, and the hantom model with the state
$\phi_1^2<\phi_2^2$. We discuss the evolution of these models in
the $\omega$-$\omega'$plane ($\omega$ is the state equation of the
dark energy, and $\omega'$ is its time derivative in unites of
Hubble time), and find that according to $\omega>-1$ or $<-1$, and
the potential of the quintom being climbed up or rolled down, the
$\omega$-$\omega'$ plane can be divided into four parts. The late
time attractor solution, if existing, is always quintessence-like
or $\Lambda$-like for hessence field, so the Big Rip doesn't
exist.  But for hantom field, its late time attractor solution can
be phantom-like or $\Lambda$-like, and sometimes, the Big Rip is
unavoidable. Then we consider two special cases: one is the
hessence field with an exponential potential, and the other is
with a power law potential. We investigate their evolution in the
$\omega$-$\omega'$ plane. We also develop a theoretical method of
constructing the hessence potential function directly from the
effective equation of state function $\omega(z)$. We apply our
method to five kinds of parametrizations of equation of state
parameter, where $\omega$ crossing $-1$ can exist, and find they
all can be realized. At last, we discuss the evolution of the
perturbations of the quintom field, and find the perturbations of
the quintom $\delta_Q$ and the metric $\Phi$ are all finite even
if at the state of $\omega=-1$ and $\omega'\neq0$.
 }
\end{quote}



e-mail: wzhao7@mail.ustc.edu.cn

\newpage
\baselineskip=19truept

\begin{center}
\Large 1. Introduction
\end{center}
Recent observations on the Type Ia Supernova (SNIa) \cite{sn},
Cosmic Microwave Background Radiation (CMB) \cite{map} and Large
Scale Structure (LSS) \cite{sdss} all suggest that the Universe
mainly consists of dark energy (73\%), dark matter (23\%) and
baryon matter (4\%). How to understand the physics of the dark
energy is an important issue, having the equation-of-state (EoS)
$\omega<-1/3$ and causing the recent accelerating expansion of the
Universe. Several scenarios have been put forward as a possible
explanation of it. A positive cosmological constant is the
simplest candidate, but it needs the extreme fine tuning to
account for the observed accelerating expansion of the Universe.
As the alternative to the cosmological constant, a lot of dynamic
models have been proposed, such as the quintessence models
\cite{quint}, which assume the dark energy is made of a light
scalar field. These models can naturally get a state of
$-1\leq\omega\leq0$, but the state of $\omega<-1$ can't be
realized, which makes many other possibilities have been
considered such as the k-essence models \cite{k} and the phantom
models \cite{phantom}, which have the non-standard kinetic terms
\cite{k}. Besides these, some other models such as the generalized
Chaplygin gas (GCG) models \cite{GCG}, the vector field models
\cite{vec,YM} also be studied by a lot of authors. Although these
models achieve some success, some problems also exist.

One essential to understand the nature of the dark energy is to
detect the value and evolution of its EoS. The observational data
shows that the cosmological constant is a good candidate
\cite{seljak}, which has the effective EoS of $\omega\equiv-1$.
However, there are several evidences showing that the dark energy
might evolve from $\omega>-1$ in the past to $\omega<-1$ today,
and cross the critical state of $\omega=-1$ in the intermediate
redshift \cite{trans}. If such a result holds on with accumulation
of observational data, this would be a great challenge to the
current models of dark energy. It is obvious that the cosmological
constant as a candidate will be excluded, and the dark energy must
be dynamical. But the normal models such as the quintessence
models, only give the state of $-1\leq\omega\leq0$. Although the
k-essence models and the phantom models can get the state of
$\omega<-1$, the behavior of $\omega$ crossing $-1$ can not be
realized \cite{vikman}. So a lot of more complex models have been
suggested to get around this \cite{models}. Obviously, the most
natural way is to consider a model with two real scalar fields. A
lot of people have studied the so-called quintom model
\cite{quintom,guo}, which is a hybrid of quintessence and phantom
(so the name quintom). Naively, We consider the action
 \be
 S=\int d^4x\sqrt{-g}\left(-\frac{\cal R}{16\pi G}+\L_{de}+\L_m\right),
 \ee
where $g$ is the determinant of the metric $g_{\mu\nu}$, $\cal R$
is the Ricci scalar, $\L_{de}$ and $\L_m$ are the lagrangian
densities of the dark energy and matter, respectively. The quintom
dark energy has the lagrangian density
 \be
 \L_{de}=\L_{Q}=\frac{1}{2}(\partial_{\mu}\phi_1)^2-\frac{1}{2}
 (\partial_{\mu}\phi_2)^2-V(\phi_1,\phi_2)~,
 \ee
where $\phi_1$ and $\phi_2$ are two real scalar fields and play
the roles of quintessence and phantom field, respectively.
Considering a spatially Flat-Robertson-Walker (FRW) Universe and
assuming the scalar fields $\phi_1$ and $\phi_2$ are homogeneous,
one obtains the effective pressure and energy density of the
quintom field
 \be
 p_Q=\frac{1}{2}\dot{\phi_1^2}-\frac{1}{2}\dot{\phi_2^2}-V(\phi_1,\phi_2)~,
 \ee
 \be
 \rho_Q=\frac{1}{2}\dot{\phi_1^2}-\frac{1}{2}\dot{\phi_2^2}+V(\phi_1,\phi_2)~,
 \ee
and the corresponding effective EoS is
 \be
 \omega_Q=\frac{\dot{\phi_1^2}-\dot{\phi_2^2}-2V(\phi_1,\phi_2)}
 {\dot{\phi_1^2}-\dot{\phi_2^2}+2V(\phi_1,\phi_2)}~.
 \ee
It is easily seen that $\omega_Q>-1$ when
$\dot{\phi_1^2}>\dot{\phi_2^2}$ is satisfied, while $\omega_Q<-1$
when $\dot{\phi_1^2}<\dot{\phi_2^2}$ is satisfied. It is obvious
that the quintom is the simplest phenomenological model of the
dark energy with $\omega_Q$ crossing $-1$. The hybrid of $\phi_1$
and $\phi_2$ in the potential function makes the models being
varied and complex, which prevents one from analyzing their
general properties. So it is interesting to look for some kinds of
quintom models with simple potentials. The cosmological evolution
of the quintom model without direct coupling between $\phi_1$ and
$\phi_2$ was studied in Ref. \cite{guo}. They showed that the
transition from $\omega_Q>-1$ to $\omega_Q<-1$ or vice versa is
possible in this type of models. But they also found that the late
attractor solutions of these quintom fields are always
phantom-like or $\Lambda$-like, which may lead to the Big Rip. The
reason is simple: since the quintessence and phantom fields
haven't direct coupling, the energy density of quintessence field
(with the EoS $\omega\geq-1$) decreases with time, but which is
increasing for phantom field (with the EoS $\omega\leq-1$). So at
last, the phantom field must be the dominant component, which may
lead to the Big Rip.

In this paper, we investigate another kind of quintom models with
the potentials
 \be\label{V}
 V(\phi_1,\phi_2)=V(\phi_1^2-\phi_2^2)~.
 \ee
In this kind of models, the fields $\phi_1$ and $\phi_2$ couple by
this potential function. Compared with the models in Ref.
\cite{guo}, these models are easily to discuss for their simple
potentials. In Ref. \cite{brane}, the authors found that this kind
of models may be the local effective approximation of the D3-brane
Universe. It is easily found that this model is equivalent with
the dark energy made of a non-canonical complex scalar field
$\Phi=\phi_1+i\phi_2$ in form with the lagrangian density
 \be\label{lag}
 \L_{de}=\frac{1}{4}\left[(\partial_{\mu}\Phi)^2
 +(\partial_{\mu}\Phi^*)^2\right]-V(\Phi^2+\Phi^{*2})~,
 \ee
which has been advised by Wei \emph{et al.} in Ref. \cite{hao},
where the authors found that this model can easily realize a state
crossing the cosmological constant boundary. It is interesting
that this model can avoid the difficulty of the Q-ball formation
which gives trouble to the spintessence. Furthermore, by choosing
a proper potential, this model can be described by a Chaplygin gas
at late time. The authors also found that the Big Rip is avoided
in the models with the exponential potential and the (inverse)
power law potential in the special cases with $\phi_1^2>\phi_2^2$.

The main task of this work is to investigate the general
characters of this kind of quintom models with the potentials in
Eq.~(\ref{V}). From the invariance under the transformation with
hyperbolic function, we separate these models into two kinds: the
hessence models with $\phi_1^2>\phi_2^2$ and hantom models with
$\phi_1^2<\phi_2^2$. By analyzing their evolution in the
$\omega$-$\omega'$ plane, we find that if $\dot{V}$ is positive
 (negative), $\omega'+3(1+\omega)(1-\omega)<0~(>0)$ is satisfied.
So the potential being climbed up or rolled down can be
immediately judged by the value of the function
$\omega'+3(1+\omega)(1-\omega)$. We also find that the hessence
field always has a quintessence-like or $\Lambda$-like attractor
solution, and the Big Rip is naturally avoided; but the hantom
field always has phantom-like or $\Lambda$-like attractor
solution, which may lead to the Big Rip. These characters can be
seen clearly in two kinds of hessence models which we have
investigated in this paper. After these, we study how to construct
the potential of hessence directly from the effective EoS:
$\omega(z)$. We apply our method to five kinds of parametrizations
of the EoS parameter, where $\omega$ crossing $-1$ can exist, and
find they all can be easily realized in the hessence models. In
the last part of this paper, we mainly discuss the evolution of
perturbations of the quintom fields. By altering the forms of the
evolutive equations, we find the divergence doesn't exist in these
equations even if at the state of $\omega=-1$ and $\omega'\neq0$.
So the values of the perturbations are finite.

The plan of this paper is as follows: in section $2$, we review
the evolutive equations of the quintom models, and separate them
into two kinds: the hessence and the hantom models. In section
$3$, we investigate their evolution in the $\omega$-$\omega'$
plane and analyze the general characters of their attractor
solutions. In section $4$, we focus on two kinds of hessence
models: one with the exponential potential and the other with the
power law potential, and study their evolution in the
$\omega$-$\omega'$ plane. In section $5$, we discuss the method to
construct the potential of hessence field directly from the
parameterized EoS and apply it to five kinds of parameterizations.
In section $6$, we investigate the perturbations of the quintom
fields and their evolutive equations. At last, in section $7$, we
have a conclusion.

We use the units $\hbar=c=1$ and adopt the metric convention as
$(+,-,-,-)$ throughout this paper.

~

\begin{center}
\Large 2. The Hessence and Hantom Models
\end{center}
The quintom field here we consider has the lagrangian density as
 \be\label{8}
 \L_{Q}=\frac{1}{2}(\partial_{\mu}\phi_1)^2-\frac{1}{2}
 (\partial_{\mu}\phi_2)^2-V(\phi_1^2-\phi_2^2)~.
 \ee
One can easily find that this lagrangian is invariant under the
transformation
 \be
 \phi_1\rightarrow\phi_1\cosh(i\alpha)-\phi_2\sinh(i\alpha)~,
 \ee
 \be
 \phi_2\rightarrow-\phi_1\sinh(i\alpha)+\phi_2\cosh(i\alpha)~,
 \ee
where $\alpha$ is constant. This property makes one can rewrite
the lagrangian density (\ref{8}) in another form
 \be\label{he}
 \L_{Q}=\L_{he}=\frac{1}{2}
 \left[(\partial_{\mu}\phi)^2-\phi^2(\partial_{\mu}\theta)^2\right]-V(\phi)~,
 \ee
where we have introduced two new variables $(\phi,~\theta)$, i.e.
 \be
 \phi_1=\phi\cosh\theta~,~~~~~~\phi_2=\phi\sinh\theta~,
 \ee
which are defined by
 \be
 \phi^2=\phi_1^2-\phi_2^2~,~~~~~~\coth\theta=\phi_1/\phi_2~.
 \ee
These models are dubbed the \emph{``hessence"} in Ref. \cite{hao}.
But it is clear that this form requires an additional requirement,
$\phi_1^2>\phi_2^2$, on the quintom models. In another condition
with $\phi_1^2<\phi_2^2$, one can rewrite the lagrangian density
in Eq.~(\ref{8}) in another form
 \be\label{ha}
 \L_Q=\L_{ha}=\frac{1}{2}\left[-(\partial_{\mu}\phi)^2
 +\phi^2(\partial_{\mu}\theta)^2\right]-V(\phi)~,
 \ee
here the variables $(\phi,~\theta)$ are defined by
 \be
 \phi^2=-\phi_1^2+\phi_2^2~,~~~~~~\coth\theta=\phi_2/\phi_1~.
 \ee
In this paper, we dub them \emph{``hantom"}. In the following
discussion, we will find that the hessence and hantom have
different properties, especially the late time attractor
solutions.

\textbf{Hessence Models}

Let us restart our discussion from the action
 \be\label{s}
 S=\int
 d^{4}x\sqrt{-g}\left(-\frac{\cal R}{16\pi G}+{\cal L}_{he}+{\cal
 L}_m\right)~,
 \ee
where the lagrangian density of hessence field can be found in
Eq.~(\ref{he}). Considering a spatially flat FRW Universe with
metric
 \be
 ds^2=dt^2-a^2 (t) \gamma_{ij}dx^idx^j~,
 \ee
where $a(t)$ is the scale factor, and $\gamma_{ij}=\delta^i_j$
denotes the flat background space. Assuming $\phi$ and $\theta$
are homogeneous, from Eqs.~(\ref{s}) and (\ref{he}), we obtain the
equations of motion for $\phi$ and $\theta$
 \be\label{1}
 \ddot{\phi}+3H\dot{\phi}+\phi\dot{\theta}^2+dV/d\phi=0~,
 \ee
 \be\label{2}
 \phi^2\ddot{\theta}+(2\phi\dot{\phi}+3H\phi^2)\dot{\theta}=0~,
 \ee
where $H\equiv\dot{a}/a$ is the Hubble parameter, an overdot
denotes the derivatives with respect to cosmic time. The pressure
and energy density of hessence field are
 \be\label{3}
 p_{he}=\frac{1}{2}\left(\dot{\phi}^2-\phi^2
 \dot{\theta}^2\right)-V(\phi)~,~~~~~~
 \rho_{he}=\frac{1}{2}\left(\dot{\phi}^2-\phi^2
 \dot{\theta}^2\right)+V(\phi)~,
 \ee
respectively. Eq.~(\ref{2}) implies
 \be
 Q=a^3\phi^2\dot{\theta}=const~,
 \ee
which is associated with the total conserved charge within the
physical volume due to the internal symmetry \cite{hao}, if
consider the hessence as a non-canonical complex scalar field. It
turns out
 \be
 \dot{\theta}=\frac{Q}{a^3\phi^2}~.
 \ee
Substituting this into Eq.~(\ref{1}), we can rewrite the kinetic
equation as
 \be\label{kinetic}
 \ddot{\phi}+3H\dot{\phi}+\frac{Q^2}{a^6\phi^3}+\frac{dV}{d\phi}=0~,
 \ee
which is equivalent to the energy conservation equation of the
hessence $\dot{\rho}_{he}+3H(\rho_{he}+p_{he})=0$. The pressure,
energy density and the EoS of the hessence are
 \be\label{24}
 p_{he}=\frac{1}{2}\dot{\phi}^2-\frac{Q^2}{2a^6
 \phi^2}-V(\phi)~,~~~~~~~
 \rho_{he}=\frac{1}{2}\dot{\phi}^2-\frac{Q^2}{2a^6 \phi^2}+V(\phi)~,
 \ee
 \be\label{25}
 \omega_{he}=\left.\left[\frac{1}{2}\dot{\phi}^2-\frac{Q^2}{2a^6
 \phi^2}-V(\phi)\right]\right/
 \left[\frac{1}{2}\dot{\phi}^2-\frac{Q^2}{2a^6
 \phi^2}+V(\phi)\right]~,
 \ee
respectively. It is easily seen that $\omega_{he}\geq-1$ when
$\dot{\phi^2}\geq Q^2/(a^6\phi^2)$, while $\omega_{he}\leq-1$ when
$\dot{\phi^2}\leq Q^2/(a^6\phi^2)$. The transition occurs when
$\dot{\phi^2}=Q^2/(a^6\phi^2)$. In the case of $Q\equiv0$, the
hessence becomes the quintessence model. If we define the
effective potential
 \be
 V_{eff}\equiv V-\frac{Q^2}{2a^6\phi^2}~,
 \ee
the kinetic equation (\ref{kinetic}) becomes
 \be
 \ddot{\phi}+3H\dot{\phi}+dV_{eff}/d\phi=0~.
 \ee
This is exactly the Klein-Gordon equation of quintessence field
with the potential $V(\phi)\equiv V_{eff}(\phi)$. So the field
$\phi$ will seek to roll towards in the minimum of its effective
potential $V_{eff}$, but which doesn't mean that $\phi$ will tend
to roll towards in the minimum of its real potential $V$. This is
the most important difference from the quintessence model. Then
when the field rolls down to its potential, when it climbs up the
potential, and how it influences the EoS of the hessence model?
This is the main task of section $3$.

\textbf{Hantom Models}

Now let's return to another case with $\phi_1^2<\phi_2^2$. The
action is
 \be\label{s2}
  S=\int
 d^{4}x\sqrt{-g}\left(-\frac{\cal R}{16\pi G}+{\cal L}_{ha}+{\cal
 L}_m\right)~,
 \ee
where the lagrangian density of the hantom can be seen in
Eq.~(\ref{ha}), which follows the kinetic equations
 \be
 \ddot{\phi}+3H\dot{\phi}+\frac{Q^2}{a^6\phi^3}-\frac{dV}{d\phi}=0~,
 \ee
where $Q=a^3\phi^2\dot{\theta}$, is the conserved charge. The
pressure, energy density and EoS are
 \ba &\disp
 p_{ha}=-\frac{1}{2}\dot{\phi}^2+\frac{Q^2}{2a^6
 \phi^2}-V(\phi)~,~~~~~~~
 \rho_{ha}=-\frac{1}{2}\dot{\phi}^2+\frac{Q^2}{2a^6 \phi^2}+V(\phi)~,\nonumber\\
 &\disp \omega_{ha}=\left.\left[-\frac{1}{2}\dot{\phi}^2+\frac{Q^2}{2a^6
 \phi^2}-V(\phi)\right]\right/
 \left[-\frac{1}{2}\dot{\phi}^2+\frac{Q^2}{2a^6
 \phi^2}+V(\phi)\right]~, \ea
respectively. It is easily seen that $\omega_{ha}\geq-1$ when
$\dot{\phi^2}\leq Q^2/(a^6\phi^2)$, and $\omega_{ha}\leq-1$ when
$\dot{\phi^2}\geq Q^2/(a^6\phi^2)$, which is inverse to the
hessence models. In the case of $Q=0$, the hantom becomes the
phantom field, which is also the origin of its name. If we define
the effective potential
 \be
 V_{eff}\equiv V+\frac{Q^2}{2a^6\phi^2}~,
 \ee
the kinetic equation becomes
 \be
 \ddot{\phi}+3H\dot{\phi}-dV_{eff}/d\phi=0~,
 \ee
This is exactly the Klein-Gordon equation of phantom field with
the potential $V(\phi)\equiv V_{eff}(\phi)$. So the field $\phi$
will seek to climb up in the maximum of its effective potential
$V_{eff}$, but which doesn't mean that $\phi$ will tend to climb
up in the maximum of its real potential $V$. Also, the discussion
on the value of $\dot V$, and its relation with $\omega$ and
$\omega'$ will be shown in the following section.

There are two important characters we should notice: First, when
encountering the condition $\phi_1^2=\phi_2^2$, one must return to
the lagrangian with general form (\ref{8}), which can't be
discussed in hessence or hantom models with functions $\phi$ and
$\theta$, but here we don't discuss this condition in this paper.
Second, in hantom, if $\phi$ is replaced by $i\phi$, one finds the
hantom becomes hessence model. So the hessence is enough to
describe all kinds of quintom fields with potential form
$V(\phi_1^2-\phi_2^2)$, if the value of $\phi$ being an imaginary
number is allowed.

~

\begin{center}
\Large 3. The Evolution of EoS of the Quintom Fields
\end{center}
Important observables to reveal the nature of dark energy are the
EoS $\omega$ and its time derivative in units of Hubble time
$\omega'$. The simplest model, cosmological constant, has the
effective state of $\omega=-1$ and $\omega'=0$, which corresponds
to a fixed point in the $\omega$-$\omega'$ plane. Generally, the
dynamics model of dark energy shows a line in this plane, which
describes the evolution of its EoS. Recently, it is shown that the
simple scalar field models of dark energy occupy rather narrow
regions in $\omega$-$\omega'$ plane \cite{oop,take}: the
quintessence has the state of $\omega\geq-1$, which only occupies
the region of $\omega'>-3(1-\omega)(1+\omega)$, and if the
quintessence has the tracker behavior, the region decreases to be
$\omega'>-(1-\omega)(1+\omega)$. A basic physics distinction in
scalar field physics requires the precision on the dynamics to be
of order $\sigma(\omega')\sim2(1+\omega)\leq0.1$ \cite{oop}. The
phantom field ($\omega\leq-1$) occupies the region of
$\omega'<-3(1-\omega)(1+\omega)$,  and if the phantom has a
tracker solution, the bound becomes
$\omega'<3\omega(1-\omega)(1+\omega)$. For the general k-essence
model with the tracker behavior, the bound on $\omega'$ is
$\omega'>\frac{3\omega}{1-2\omega}(1-\omega)(1+\omega)$.
%
%
%
To confirm the quintom models, the crossing of cosmological
constant must be found. So the dynamics of dark energy, especially
at high redshift, is very important. The SNAP mission is expected
to observe about $2000$ SNIa each year, over a period of three
years. Most of these SNIa are at the redshift $z\in[0.2, ~1.2]$.
The SNIa plus weak lensing methods conjoined can determine the
present equation of state ratio, $\omega_0$, to $5\%$, and its
time variation, $\omega'$, to $0.11$ \cite{snap}. It has a
powerful ability to differentiate the various dark energy models.

In this section, we will extend the phase space analysis to the
quintom fields, where $\omega$ crossing $-1$ exists. First we
consider the \emph{\textbf{hessence}} model, which has the kinetic
equation
 \be\label{28}
 \ddot{\phi}+3H\dot{\phi}+\frac{Q^2}{a^6\phi^3}+\frac{dV}{d\phi}=0~.
 \ee
If $\omega\neq-1$ is satisfied, one can define a function
 \be
 x\equiv\left|\frac{1+\omega}{1-\omega}\right|=\left|\frac{\frac{1}{2}
 \dot{\phi}^2-\frac{Q^2}{2a^6\phi^2}}{V}\right|~.
 \ee
Then the kinetic equation (\ref{28}) follows that
 \be\label{32}
 1+\frac{1}{6}\frac{d\ln x}{d\ln
 a}=-\frac{1}{3HV}\frac{\dot{V}}{1+\omega}~.
 \ee
where $a$ is the scale factor, and we have set the present scalar
factor $a_0=1$. In the case of $Q\equiv0$, hessence becoming the
normal quintessence field, the formula~(\ref{32}) leads to the
relation \cite{oop}
 \be
 \mp\frac{V'}{V}=\sqrt{\frac{3\kappa^2(1+\omega)}
 {\Omega_{\phi}}}\left(1+\frac{1}{6}\frac{d\ln x}{d\ln a}\right)~,
 \ee
where $\kappa^2=8\pi G$, and $\Omega_{\phi}$ is the energy density
of the quintessence field. The minus sign corresponds to
$\dot{\phi}>0~(V'<0)$ and plus sign to the opposite. This can
follow a constraint on $\omega'$,
 \be\label{34}
 \omega'>-3(1-\omega)(1+\omega)~,
 \ee
where and before $\omega'=d\omega/d\ln a$. This bound applies to a
general class of quintessence field which monotonically rolls down
the potential.

Here we return to the general case of the formula (\ref{32}) with
$Q\neq0$. Define a useful function
$c_a^2\equiv\dot{p}/\dot{\rho}$. If the matter is a kind of
prefect liquid, this function is the adiabatic sound speed of this
liquid. But for the scalar field, this function isn't a real
speed. For the hessence field, it can be written as
 \be\label{35}
 c_a^2=\frac{2\dot{V}}{3H(1+\omega)\rho}+1~,
 \ee
where $\rho$ is the energy density of hessence, and $\omega$ is
its EoS. If the hessence returns to the quintessence field, one
always has $c_a^2<1$, which is for the field $\phi$ monotonically
rolls down $(\dot{V}<0)$ the potential in the quintessence model
with EoS $\omega>-1$. And for phantom field with $\omega<-1$,
$c_a^2<1$ is also satisfied, since the phantom field climbs up the
potential $(\dot{V}>0)$. But here for the hessence model,
$c_a^2>1$ can exist, which only needs $\dot{V}(1+\omega)>1$. In
these kinds of models, the function $c_a$ isn't a physical speed.
So the case of $c_a^2>1$ is consistent with the Relativity Theory.
Here this function reflects the evolutive direction of the
potential of the hessence. In the case of $\omega>-1$, i.e. the
quintessence-like, $c_a^2>1~(c_a^2<1)$ indicates that $\dot{V}>0~
(\dot{V}<0)$, the field $\phi$ climbing up (rolling down) its
potential, and in the case of $\omega<-1$, i.e. the phantom-like,
$c_a^2>1 ~(c_a^2<1)$ indicates that $\dot{V}<0 ~(\dot{V}>0)$, the
field $\phi$ rolling down (climbing up) its potential. Inserting
this function into Eq.~(\ref{32}), one gets
 \be\label{1+6}
 1+\frac{1}{6}\frac{d\ln x}{d\ln
 a}=\frac{1-c_a^2}{2V}\rho~,
 \ee
which can be rewritten as
 \be\label{37}\left.
 \left(1+\frac{1}{6}\frac{d\ln x}{d\ln
 a}\right)\right/\left(1-c_a^2\right)=\frac{\rho}{2V}>0~.
 \ee
Using the relation of
 \be
 \frac{d\ln x}{d\ln
 a}=\frac{2\omega'}{(1+\omega)(1-\omega)}~,
 \ee
Eq.~(\ref{37}) follows that
 \be
 \frac{\omega'}{(1+\omega)(1-\omega)(1-c_a^2)}>\frac{-3}{(1-c_a^2)}~.
 \ee
So the $\omega$-$\omega'$ plane is divided into four parts

 ~~~~~~~~~~~~~~~~~~~~~~~~~~~~~~~\emph{I}:~~~~~~ $c_a^2<1$ ~~\&~~
 $\omega>-1$,~~~~~$\omega'>-3(1-\omega)(1+\omega)$~;

  ~~~~~~~~~~~~~~~~~~~~~~~~~~~~~~\emph{II}:~~~~~~~$c_a^2>1$ ~~\&~~
 $\omega<-1$,~~~~~$\omega'>-3(1-\omega)(1+\omega)$~;

  ~~~~~~~~~~~~~~~~~~~~~~~~~~~~~\emph{III}:~~~~~~ $c_a^2<1$ ~~\&~~
 $\omega<-1$,~~~~~$\omega'<-3(1-\omega)(1+\omega)$~;

  ~~~~~~~~~~~~~~~~~~~~~~~~~~~~~\emph{IV}:~~~~~~ $c_a^2>1$ ~~\&~~
 $\omega>-1$,~~~~~$\omega'<-3(1-\omega)(1+\omega)$~.\\
This can be seen clearly in Fig.~[1]. From Eq.~(\ref{35}), one can
easily find that $\dot{V}<0$ is satisfied in Region \emph{I} and
\emph{II}, the field rolling down the potential, and $\dot{V}>0$
is satisfied in Region \emph{III} and \emph{IV}, the field
climbing up the potential. So from the value of the function
$\omega'+3(1-\omega)(1+\omega)$ being positive or negative, one
can immediately judge how the field evolves at that time. This is
one of the most important results in this section. In the case of
$Q=0$, the hessence returns to the quintessence field, and
conditions of $c_a^2<1$ and $\omega>-1$ are always satisfied. So
only Region \emph{I} is allowed, and the bound of
$\omega'>-3(1-\omega)(1+\omega)$ is satisfied, which is exactly
same with Eq.~(\ref{34}). In the general case of $Q\neq0$, these
four regions are all allowed.

Now, let's focus on the issue: how the EoS crosses $-1$ in the
$\omega$-$\omega'$ plane? Assuming at some time, the hessence
being in Region \emph{I} with $\omega>-1$ and $c_a^2<1$, there are
two ways for the field to run to the regions with $\omega<-1$:

a) One is that, the field runs across the \textbf{Critical}
\textbf{Point} $(\omega,\omega')=(-1,0)$, and arrives in Region
\emph{II} or \emph{III}. Unfortunately, this can't be realized in
finite time. We use Taylor expansion of $\omega'$ at the state
around the Critical Point, and keep the first two terms,
 \[
 \omega'=\left.\omega'\right|_{-1}+\left.\frac{\partial\omega'}
 {\partial\omega}\right|_{-1}(\omega+1)\equiv
 b(\omega+1)~,
 \]
where $b$ is a constant number. Using the definition of $\omega'$,
this equation yields that $|\omega+1|=a^b$. It is possible to get
$\omega=-1$ at $a\neq0$, only if $b<0$ is satisfied. In this
condition, only if $a\rightarrow\infty$, $\omega=-1$ can be got.
So the hessence field can't cross the Critical Point in finite
time.

b) The other way is to cross the dot line and arrive in Region
\emph{II}. This is the only way for the hessence field to cross
the state of $\omega=-1$. From Eq.~(\ref{35}), one can easily get
 \be
 c_a^2=\omega-\frac{\omega'}{3(1+\omega)}~.
 \ee
When $\omega=-1$ and $\omega'\neq0$, ~$c_a^2$ is divergent. In
section $6$, we will prove that this divergence doesn't yield the
physical divergence of the perturbations of the dark energy.

%

At last, we discuss the possible late time attractor solutions of
the hessence field. There are three kinds of solutions:~ a)~~The
hessence has not an oscillating EoS and the late time attractor is
phantom-like with $\omega<-1$. Since $\omega'=0$ is satisfied for
the attractor, this solution must be in the Region \emph{III};~
b)~~For the similar reason, if the attractor is quintessence-like
with $\omega>-1$, it must stay in Region \emph{I};~ c)~~The other
possibility is the $\Lambda$-like attractor. In this case, the
hessence will run to the Critical Point $(\omega,\omega')=(-1,0)$.
But the phantom-like attractor is difficult to realize. From the
expressions of the pressure $p$ and the energy density $\rho$ of
the hessence, one knows that only if
$\frac{1}{2}\dot{\phi}^2<\frac{Q^2}{2a^6\phi^2}$, the attractor is
phantom-like. If $\dot{|\phi|}>0$ is satisfied for the attractor,
the value of $\frac{Q^2}{2a^6\phi^2}$ will damp quickly with time,
and at last, it is always unavoidable to arrive at
$\frac{1}{2}\dot{\phi}^2>\frac{Q^2}{2a^6\phi^2}$, which is the
quintessence-like result; on the other hand, if $\dot{|\phi|}<0$
is satisfied for the attractor, it is inevitable to run to the
state of $|\phi|=0$, which is forbidden by the definition of the
hessence model. So in the hessence models, the Big Rip is avoided
naturally, which has been discussed in some special examples in
Ref. \cite{hao}.

Now, let's discuss the \emph{\textbf{hantom}} model in the similar
way. The pressure and energy density of the hantom are
 \be
 p=-\frac{1}{2}\dot{\phi}^2+\frac{Q^2}{2a^6\phi^2}-V~,~~~
 \rho=-\frac{1}{2}\dot{\phi}^2+\frac{Q^2}{2a^6\phi^2}+V~,
 \ee
respectively. And the state equation is
 \be
 \omega=\left.\left[\frac{1}{2}\dot{\phi}^2-\frac{Q^2}{2a^6\phi^2}+V\right]\right/
 \left[\frac{1}{2}\dot{\phi}^2-\frac{Q^2}{2a^6\phi^2}-V\right]~.
 \ee
The kinetic equation is
 \be
 \ddot{\phi}+3H\dot{\phi}+\frac{Q^2}{a^6\phi^3}-\frac{dV}{d\phi}=0~,
 \ee
which follows that
 \be
 \frac{\omega'}{(1+\omega)(1-\omega)(1-c_a^2)}>\frac{-3}{1-c_a^2}~,
 \ee
where $c_a^2$ is also defined by $c_a^2\equiv\dot{p}/\dot{\rho}$,
and the relation (\ref{35}) is also satisfied. So in the hantom
models, we also can divide the $\omega$-$\omega'$ plane into the
exactly same four parts as in Fig.~[1]. One can easily find that
$\dot{V}<0$ is satisfied in Region \emph{I} and \emph{II}, the
field rolling down the potential, and $\dot{V}>0$ is satisfied in
Region \emph{III} and \emph{IV}, the field climbing up the
potential. In the case of $Q=0$, the hantom returns to the phantom
field, and the conditions of $c_a^2<1$ and $\omega<-1$ are always
satisfied. So only Region \emph{III} is allowed, and the bound of
$\omega'<-3(1-\omega)(1+\omega)$ is satisfied. For the similar
reason as before, the late time attractor of hantom can be
phantom-like (Region \emph{III}) or $\Lambda$-like (Critical
Point). The former can lead to the Big Rip at the late Universe.
In the following sections, we only discuss the hessence models to
avoid the Big Rip.

~

\begin{center}
\Large 4. Two kinds of Hessence Models
\end{center}
In this section, we discuss two kinds of special potentials of the
hessence fields. One is the model with an exponential potential
 \be
 V(\phi)=V_0 e^{-\lambda\kappa\phi}~,
 \ee
where $\lambda$ is a dimensionless constant. The other is the
model with a power law potential
 \be
 V(\phi)=V_0(\kappa\phi)^n~,
 \ee
where $n$ is a dimensionless positive constant. These two forms of
potentials are the most popular models, which are discussed in the
scalar dark energy. In this section, we will numerically solve the
kinetic equation of the hessence with these two kinds of potential
functions, and study the evolution of $\omega$ and $\omega'$ in
detail to check the results we mentioned before. We focus on four
special models:

Model \emph{a1}: $\dot{\phi_{0}}>0$,~$V(\phi)=V_0
e^{-\lambda\kappa\phi}$ with
$\lambda=1.0$,~$Q^2/(\rho_t\phi_0^2)=5$, ~$\omega_{0}=-1.4$~;

Model \emph{a2}: $\dot{\phi_{0}}<0$,~$V(\phi)=V_0
e^{-\lambda\kappa\phi}$ with
$\lambda=1.0$,~$Q^2/(\rho_t\phi_0^2)=0.5$, ~$\omega_{0}=-0.7$~;

Model \emph{b1}: $\dot{\phi_{0}}>0$,~$V(\phi)=V_0 (\kappa\phi)^n$
with $n=2$,~$Q^2/(\rho_t\phi_0^2)=5$, ~$\omega_{0}=-1.4$~;

Model \emph{b2}: $\dot{\phi_{0}}<0$,~$V(\phi)=V_0 (\kappa\phi)^n$
with
$n=2$,~$Q^2/(\rho_t\phi_0^2)=0.5$, ~$\omega_{0}=-0.7$~,\\
where $\phi_0$ is the field $\phi$ with the present value,
$\rho_t$ is the present total energy density, and $\omega_{0}$ is
the present EoS of the hessence. In all these models, we choose
the present density parameters $\Omega_{he0}=0.7$ and
$\Omega_{m0}=0.3$. The first two models have the exponential
potentials, and the latter two ones have power law potentials. The
present EoS in Models \emph{a1} and \emph{b1} are phantom-like,
and which are quintessence-like in Models \emph{a2} and \emph{b2}.
The EoS of the hessence is
 \be
 \omega=\left.\left[\frac{1}{2}\dot{\phi}^2-\frac{Q^2}{2a^6
 \phi^2}-V(\phi)\right]\right/
 \left[\frac{1}{2}\dot{\phi}^2-\frac{Q^2}{2a^6
 \phi^2}+V(\phi)\right]~.
 \ee
In Fig.~[2], we plot their evolution in $\omega$-$\omega'$ plane,
and find that, except the Model \emph{b2}, the behavior of
$\omega$ crossing $-1$ exists in all these models. The Models
\emph{a1} and \emph{a2} run to the same attractor solution of
$(\omega,\omega')=(-2/3,0)$, i.e. the quintessence-like solution,
and the Models \emph{b1} and \emph{b2} run to the same point of
$(\omega,\omega')=(-1,0)$, i.e. the $\Lambda$-like solution. These
results are same with the conclusion in Ref. \cite{hao}, where the
authors found that the hessence models with exponential potentials
have the stable attractor solutions with $\omega=-1+\lambda^2/3$,
and the models with power law potentials have the stable attractor
solutions with $\omega=-1$. In these models, EoS crossing $-1$
obeys the second way: crossing the dot line (excluding the
Critical Point $(\omega,\omega')=(-1,0)$). So the divergence of
the function $c_a^2$ exists.

In the $\omega$-$\omega'$ plane, the Models \emph{a1} and
\emph{b2} stay in the region \emph{I} and \emph{II} at all times,
so the condition $\dot{V}<0$ holds for all time, and the fields
roll down their potentials. But for the Models \emph{b1} and
\emph{a2}, they run from the regions with
$\omega'<-3(1+\omega)(1-\omega)$ to the ones with
$\omega'>-3(1+\omega)(1-\omega)$, so the fields climb up at the
beginning and then roll down the potentials. These can be seen
clearly in Fig.~[3], where we plot the evolution of function
$f\equiv\dot{V}/H\rho$ with the scale factor.

~

\begin{center}
\Large 5. Construct the Potentials of the Hessence Fields
\end{center}
Generally, the observed EoS of dark energy is a function of
redshift $z$, and the function form of $\omega(z)$ depends on the
parametrized model. Now, how to know the potential function from
the observed $\omega(z)$? If realized, it will be a direct way to
relate the observation and dark energy models. In Ref.
\cite{construct}, the authors suggested a theoretical method of
constructing the quintessence potential $V(\phi)$ directly from
the state function $\omega(z)$. Since $\omega<-1$ can't be
realized in the quintessence models, this method is effective only
for the state of $-1\leq\omega\leq1$. But the recent observations
mildly suggest that $\omega$ crossing $-1$ is existing. In this
section, we will develop this method to construct the hessence
potential $V(\phi)$ directly from $\omega(z)$. We apply this
method to five typical parametrizations.

Consider the FRW Universe, which is dominated by the
non-relativistic matter and a spatially homogeneous hessence field
$\phi$. The Friedmann equation is
 \be\label{52}
 H^2=\frac{\kappa^2}{3}\left(\rho_m+\rho_{he}\right)~,
 \ee
where $\rho_m$ and $\rho_{he}$ are the densities of matter and
hessence, respectively. The pressure, energy density and EoS of
the hessence field have been written in Eqs.~(\ref{24}) and
(\ref{25}), from which we have
 \be\label{55}
 V(\phi)=\frac{1}{2}\left(1-\omega_{he}\right)\rho_{he}~,
 \ee
 \be\label{56}
 \dot{\phi}^2=\frac{Q^2}{a^6\phi^2}+(1+\omega_{he})\rho_{he}~.
 \ee
These two equations relate the potential $V$ and field $\phi$ to
the only function $\rho_{he}$. So the main task below is to solve
the function form $\rho_{he}(z)$ from the parametrized EoS
$\omega_{he}(z)$. The energy conservation equation of the hessence
field is
 \be
 \dot{\rho}_{he}+3H(\rho_{he}+p_{he})=0~,
 \ee
which yields
 \be\label{58}
 \rho_{he}(z)=\rho_{he0} \exp\left[3\int_0^z(1+\omega_{he})d\ln(1+z)\right]\equiv
 \rho_{he0}E(z)~,
 \ee
where the subscript $0$ denotes the value of a quantity at the
redshift $z=0$ (present). In term of $\omega_{he}(z)$, the
potential can be written as a function of the redshift $z$:
 \be
 V[\phi(z)]=\frac{1}{2}(1-\omega_{he})\rho_{he0}E(z)~.
 \ee
With the help of $\rho_m=\rho_{m0}(1+z)^3$ and Eq.~(\ref{58}), the
Friedmann Eq.~(\ref{52}) becomes
 \be
 H(z)=H_0\left[\Omega_{m0}(1+z)^3+\Omega_{he0}E(z)\right]^{1/2}~,
 \ee
where $\Omega_{m0}$ and $\Omega_{he0}$ are the present relativity
densities of matter and hessence, respectively. Using
Eq.~(\ref{56}), we have
 \be\label{61}
 \frac{d\phi}{dz}=\mp\frac{\left[\frac{Q^2}{a^6\phi^2}+(1+\omega_{he})\rho_{he}\right]^{1/2}}{(1+z)H(z)}~,
 \ee
where the upper (lower) sign applies if $\dot{\phi}>0~
(\dot{\phi}<0)$. Here we choose the lower sign to avoid the state
of $\phi=0$. It is helpful to define three dimensionless
quantities $\tilde{\phi}$, $\tilde{V}$ and $C$
 \be\label{62}
 \tilde{\phi}\equiv\kappa\phi~,~~~~~\tilde{V}\equiv V/\rho_{he0}~,
 ~~~~~C\equiv\kappa^2Q^2/\rho_{he0}~.
 \ee
Eqs.~(\ref{55}) and (\ref{56}) become
 \be\label{63}
 \frac{d\tilde{\phi}}{dz}=\frac{\sqrt{3}}{(1+z)}
 \left[\frac{C(1+z)^6\tilde{\phi}^{-2}+(1+\omega_{he})E(z)}{r_0(1+z)^3+E(z)}\right]^{1/2}~,
 \ee
 \be\label{64}
 \tilde{V}[\phi]=\frac{1}{2}(1-\omega_{he})E(z)~,
 \ee
where $r_0\equiv\Omega_{m0}/\Omega_{he0}$ is the energy density
ratio of matter to hessence at present time.  These two equations
relate the hessence potential $V(\phi)$ to the EoS of the hessence
$\omega_{he}(z)$. Given an effective $\omega_{he}(z)$, the
construction Eqs.~(\ref{63}) and (\ref{64}) allow us to construct
the hessence potential $V(\phi)$. Here we consider the
construction process with the following five parametrization
methods.\\
The first model we consider is the EoS with constant
value \cite{constant}:\\
Model \emph{a}: $\omega_{he}=\omega_0$. If $\omega_0>-1$, a
quintessence-like value, the construction of the potential can be
easily realized with $Q\equiv0$, where the hessence returns to the
quintessence field. This condition has been discussed in Ref.
\cite{construct}. Here we consider another case with
$\omega_0=-1.2<-1$, a phantom-like value, and construct its
potential function. In this case, the function $E(z)$ has a simple
form
 \be
 E(z) = (1+z)^{3(1+w_0)}~.
 \ee
Then we consider three two-parameter models \cite{1,2,3}:\\
Model \emph{b}: $\omega_{he}=\omega_0+\omega_1 z$, and the
function $E(z)$ has the form
 \be
 E(z) = (1+z)^{3(1+w_0-w_1)}e^{3w_1 z}~,
 \ee
Model \emph{c}: $\omega_{he}=\omega_0+\omega_1 \frac{z}{1+z}$, and
the function $E(z)$ has the form
 \be
 E(z) = (1+z)^{3(1+w_0+w_1)}e^{-3w_1 \frac{z}{1+z}}~,
 \ee
Model \emph{d}: $\omega_{he}=\omega_0+\omega_1\ln(1+z)$, and the
function $E(z)$ has the form
 \be
 E(z) = (1+z)^{3(1+w_0)+\frac{3}{2}w_1\ln(1+z)}~.
 \ee
Inserting these into Eqs.~(\ref{63}) and (\ref{64}), we can
numerically evaluate the potential functions. In Fig. [4], we plot
these parametrized EoS $\omega_{he}(z)$, where we have chosen the
parameters $\omega_0=-1.2$, $\omega_1=0.5$. We find that, in the
Models \emph{b}, \emph{c} and \emph{d}, $\omega$ crossing $-1$
exists. In Fig.~[5], we plot the evolution of $\phi$ with the
redshift $z$, where we have chosen the parameters $r_0=3/7$, $C=5$
and the present value $\tilde{\phi}_0=1$. One finds, with the
increasing of redshift, the values of $\phi$ monotonically
increase in all these models. So the condition of $\phi=0$ is
avoided.

From Eq.~(\ref{64}), one can get the evolution of the potentials
$V(z)$ of the hessence fields, which have been shown in Fig. [6].
From this figure, one finds the potential is climbed up for all
time in the Model \emph{a}. And in the Models \emph{b}, \emph{c}
and \emph{d}, the fields roll down the potentials at the higher
redshift and climb up the potentials at the lower redshift. They
all arrive at the lowest points of their potentials at the
redshift, where $\omega_{he}'+3(1-\omega_{he})(1+\omega_{he})=0$
is satisfied. These results are exactly what we expect: the Model
\emph{a} with $\omega_{he}<-1$ and $\omega'_{he}=0$, stays in
Region \emph{III} in the $\omega_{he}$-$\omega_{he}'$ plane, so
$\dot{V}>0$ is satisfied for all time. But for the other models,
we plot them in the $\omega_{he}$-$\omega_{he}'$ plane in Fig.
[7]. At the lower redshift, they all stay in Region \emph{III},
which makes $\dot{V}>0$ is satisfied. And at the higher redshift,
they all arrive in the Region \emph{I}, where $\dot{V}<0$ is
satisfied. Combining Eqs.~(\ref{63}) and (\ref{64}), the potential
functions $V(\phi)$ can be got, which have been shown in Fig.[8].
One finds these potentials are not monotonic functions of $\phi$,
except the Model \emph{a}, which are obviously different from the
normal quintessence models \cite{construct}.


Recently, a lot of authors have considered the dark energy with
oscillating EoS \cite{osci}. They discussed that this kind of
models give a naturally answer for the ``coincidence problem" and
``fine-tuning problem" of the dark energy. And in some models, it
can naturally relate the early inflation and the recent
accelerating expansion. The most interesting is that these models
are likely to be marginally suggested by some observations
\cite{ob}. The difficulty is that this kind of EoS is difficult to
realize from the general potential function. Many periodic or
nonmonotonic potentials have been put forward for quintessence
fields, but rarely give rise to periodic $\omega(z)$. Here we
consider a kind of
oscillating parametrization:\\
Model \emph{e}:
$\omega_{he}=\omega_0+\omega_1\sin(\frac{1+z_c}{1+z})$. At the
high redshift $z\gg z_c$, the oscillation of $\omega_{he}(z)$
disappears, and $\omega_{he}\simeq\omega_0$. The EoS is
oscillating only when $z<z_c$. Here we choose parameters:
$\omega_0=-0.7$ $\omega_1=0.5$ and $z_c=10$, so $\omega_{he}$
crossing $-1$ exists. And the present EoS is $\omega_{he0}=-1.2$,
and $\omega_{he0}'=0.5$, which are same to the values in Models
\emph{b}, \emph{c} and \emph{d}. The observations mildly suggest
that the EoS of dark energy crossed $-1$ at very recent, which had
been regarded as the second cosmological coincidence problem. The
parametrization in Model \emph{e} gives a naturally answer for
this problem. Using this $\omega_{he}(z)$, we also can construct
the potential of the hessence by applying the Eqs.~(\ref{63}) and
(\ref{64}), which has been plotted in Figs. [6] and [8]. We find
although the potential $V(\phi)$ shows an oscillating behavior,
this oscillation is different from the simple sine or cosine
function, which appears at the pseudo-Nambu-Goldstone boson (PNGB)
field \cite{pngb} with the potential
$V(\phi)=V_0[1+\cos(\phi/f)]$, where $f$ is a (axion) symmetry
energy scale. Here the potential $V(\phi)$ of the hessence is an
oscillating function with the increasing (or decreasing)
amplitude. This suggests the method to build the potential of the
scalar field dark energy, which can yield an oscillating EoS.

~

\begin{center}
\Large 6. The Perturbations of the Quintom Fields
\end{center}
If the dark energy is a kind of dynamical field (or liquid), it is
necessary to consider the perturbations of it. These studies have
been done for many kinds of dark energy models, such as the
quintessence fields, the phantom fields, the k-essence fields, and
so on. Some models have predicted too large perturbations of the
dark energy or the background metric. For example, the GCG models
can produce the oscillations or exponential blowup of the matter
power spectrum, which is inconsistent with observations
\cite{unify}; the Yang-Mills field models have the imaginary sound
speed, which makes the perturbations of the intrinsic spatial
curvature $\Phi$ increasing rapidly at recent epoch \cite{YM}. For
many models, which allows the existence of EoS crossing $-1$, the
perturbations of the dark energy may be divergent at the state of
$\omega_{de}=-1$ \cite{ZhaoGB}. But whether or not, this
divergence exists in our quintom models? In this section, we focus
on this question by discussing the evolution of the perturbations
of our quintom fields. In the conformal Newtonian gauge, the
perturbed metric is given by
 \be\label{me}
 ds^2=a^2(\tau)\left[(1+2\Phi)d\tau^2-(1-2\Psi)dx^idx_i\right]~,
 \ee
here we have used the conformal time $\tau$, which relates to the
cosmic time by $dt\equiv ad\tau$. The gauge-invariant metric
perturbation $\Psi$ is the Newtonian potential and $\Phi$ is the
perturbation to the intrinsic spatial curvature. Always the
background matters in the Universe are perfect fluids without
anisotropic stress, which follows that $\Phi=\Psi$. So there is
only one perturbation function $\Phi$ in the metric (\ref{me}).

Using the notations of Ref. \cite{ma}, the perturbations of the
dark energy (including our quintom field) satisfy
 \be\label{delta}
 \delta_{de}'=-(1+\omega_{de})(\theta_{de}-3\Phi')-3\H(c_s^2-\omega_{de})\delta_{de}~,
 \ee
 \be\label{theta}
 \theta_{de}'=-\H(1-3\omega_{de})\theta_{de}+
3\H(c_a^2-\omega_{de})\theta_{de}
 +k^2\left(\frac{c_s^2}{1+\omega_{de}}\delta_{de}+\Phi\right)~,
 \ee
where $\H\equiv a'/a$, and the $'$prime$'$ denotes $d/d\tau$.
$c_s$ is the sound speed of the dark energy, which is defined by
$c_s^2\equiv\delta p_{de}/\delta\rho_{de}$, and its value can't be
larger than the light speed $c$. Here we have assumed zero
anisotropic stress, which is the case for matter and simple dark
energy models. The perturbation $\delta_{de}$ is defined by
$\delta_{de}\equiv\delta\rho_{de}/\rho_{de}$, and $\theta_{de}$ is
the divergence of the velocity of the dark energy. We should point
out that both scalar fields and fluids obey the same forms of
these two equations, and the only difference comes from the term
of $c_s^2$. From Eq.~(\ref{theta}), one can find that when
$c_a^2\rightarrow\infty$, where $\omega_{de}=-1$ and
$\omega_{de}'\neq0$ are satisfied, one will get infinite
$\theta_{de}'$. Fortunately, this divergence can't lead to the
divergence of $\delta_{de}$. It is helpful to define another
function
 \be
 \vartheta_{de}\equiv(1+\omega_{de})\theta_{de}~,
 \ee
then Eqs.~(\ref{delta}) and (\ref{theta}) become
 \be\label{11}
 \delta_{de}'=-\vartheta_{de}+3\Phi'(1+\omega_{de})-3\H(c_s^2-\omega_{de})\delta_{de}~,
 \ee
 \be\label{22}
 \vartheta_{de}'=-\H(1-3\omega_{de})\vartheta_{de}+k^2c_s^2\delta_{de}+k^2(1+\omega_{de})\Phi~.
 \ee
We find that the divergence at $\omega_{de}=-1$ disappears. For
the hessence field, we have
 \be
 \delta\rho_{he}=\dot{\phi}\dot{(\delta\phi)}+\frac{Q^2\delta\phi}{a^6\phi^3}+\frac{d
 V}{d\phi}\delta\phi-\Phi\dot{\phi}^2~,
 \ee
  \be
 \delta p_{he}=\dot{\phi}\dot{(\delta\phi)}+\frac{Q^2\delta\phi}{a^6\phi^3}-\frac{d
 V}{d\phi}\delta\phi-\Phi\dot{\phi}^2~.
 \ee
In the frame where the perturbations of the scalar field
$\delta\phi$ and $\dot{(\delta\phi)}$ are negligible, the sound
speed of the quintom becomes $c_s^2\simeq1$. So Eqs.~(\ref{11})
and (\ref{22}) become
 \be\label{111}
 \delta_{he}'=-\vartheta_{he}+3\Phi'(1+\omega_{he})-3\H(1-\omega_{he})\delta_{he}~,
 \ee
 \be\label{222}
 \vartheta_{he}'=-\H(1-3\omega_{he})\vartheta_{he}+k^2\delta_{he}+k^2(1+\omega_{he})\Phi~.
 \ee
In general the evolution of the perturbations can be numerically
computed, which depends on the component in the Universe and the
special quintom models. For a complete study on the perturbations,
the evolution of the metric perturbation $\Phi$ should also been
considered, which satisfies the equation \cite{evolution,ZhaoGB}
 \be\label{Phi}
 \Phi''+3\H\left(1+\frac{p_t'}{\rho_t'}\right)\Phi'+\frac{p_t'}{\rho_t'}k^2\Phi
 +\left[\left(1+3\frac{p_t'}{\rho_t'}\right)\H^2+2\H'\right]\Phi=4\pi
 Ga^2\left(\delta p_t-\frac{p_t'}{\rho_t'}\delta\rho_t\right)~.
 \ee
The pressure $p_t\equiv\sum_ip_i$, and energy density
$\rho_t\equiv\sum_i\rho_i$, which should include the contributions
of baryon, photon, neutrino, cold dark matter, and dark energy.
Especially at late time of the Universe, the effect of dark energy
is very important. Combining the Eqs.~(\ref{111}), (\ref{222}) and
(\ref{Phi}), one can numerically solve the function $\Phi(\tau)$
for special quintom models. Although we won't calculate them in
this paper, it also can be found that the value of $\Phi$ is
finite as if the total function $p_t'/\rho_t'$ is finite. So even
if $c_a^2\equiv p_{de}'/\rho_{de}'$ is divergent, $\Phi$ is also
finite if only $p_t'/\rho_t'$ isn't divergent. We should notice
that, the Eqs.~(\ref{111}), (\ref{222}) and (\ref{Phi}) are also
satisfied for the hantom models, where we only need to replace
$\delta_{he}$ with $\delta_{ha}$,  $\vartheta_{he}$ with
$\vartheta_{ha}$, and $\omega_{he}$ with $\omega_{ha}$. We remind
that the perturbations $\Phi$ and $\Psi$ can directly influence
the CMB anisotropy power spectrum by the
integral-Sachs-Wolfe~(ISW) effect \cite{sw},
 \be
 C_l^{ISW}\propto\int\frac{dk}{k}\left[\int_0^{\chi_{LSS}}d\chi~(\Phi'+\Psi')j_l(k\chi)\right]^2~,
 \ee
where $\chi_{LSS}$ is the conformal distance to the last
scattering surface and $j_l$ the $l'$th spherical Bessel function.
The ISW effect occurs because photons can gain energy as they
travel through time-varying gravitational wells. One always solves
the power spectrum $C_l$ in the numerical methods
\cite{numerical}, which can directly compare with the observations
\cite{map}. This is one of the most important way to study the
dark energy models.

~

\begin{center}
\Large 7. Conclusion
\end{center}
Understanding the nature of dark energy is one of the most
important issues in the modern cosmology. Until recently, the most
effective way is to detect the EoS $\omega_{de}$ and its time
derivative $\omega_{de}'$ by the observations on SNIa, CMB ,LSS
and so on. There are mild evidences to show that $\omega_{de}$
crossing $-1$ exists at the very low redshift, which makes the
building of the dark energy models difficult. A simple
quintessence, phantom or k-essence field is insufficient. Although
the states of $\omega_{de}\geq-1$ and $\omega_{de}\leq-1$ can be
realized in these models, they all can't give a state of
$\omega_{de}$ crossing $-1$. A lot of more complex models have
been suggested to account for this problem. Among them, the
simplest one is the quintom model, which is a hybrid of
quintessence and phantom fields. This kind of models have been
discussed by a lot of authors. In this paper, we focused on a kind
of special quintom, which has the potential
$V(\phi_1^2-\phi_2^2)$, where $\phi_1$ and $\phi_2$ are the
quintessence and phantom fields, respectively. We investigated the
general characters of this kind of models.

The lagrangian densities of the quintom are invariant under the
hypergeometric transformation between the fields $\phi_1$ and
$\phi_2$, which makes one can separate the quintom into two kinds:
the hessence and the hantom. The former has the state of
$\phi_1^2>\phi_2^2$, and the latter has the state of
$\phi_1^2<\phi_2^2$. We discussed their evolution in the
$\omega_{Q}$-$\omega_{Q}'$ plane, and found this plane can be
divided into four parts according to the values of $\dot{V}$ and
$\omega_{Q}+1$ being larger or smaller than zero. The fact
$\dot{V}>0~(<0)$ denotes that the potential of the quintom is
climbed up (rolled down), and the fact $\omega_Q>-1~(<-1)$ denotes
the field is quintessence-like (phantom-like). From their kinetic
equations, we found $\omega_Q'>-3(1-\omega_Q)(1+\omega_Q)$ is
satisfied for the case of $\dot{V}<0$, and
$\omega_Q'<-3(1-\omega_Q)(1+\omega_Q)$ is satisfied when
$\dot{V}>0$, which directly relates the evolution of potential to
the value of EoS~$\omega_Q$. We also found that, if the late time
attractor solution exists, which is always quintessence-like or
$\Lambda$-like for the hessence field, so the Big Rip is naturally
avoided. But for hantom, this solution can be phantom-like or
$\Lambda$-like. These characters are clearly shown in two hessence
models with the exponential potential and power law potential.

In this paper, We also developed a theoretical method of
constructing the hessence potential directly from the observable
EoS $\omega_{he}(z)$. We applied our method to five kinds of
parametrizations of EoS parameter, where $\omega_{he}$ crossing
$-1$ can exist, and found they all can be realized in hessence
models. Especially, the fifth model with the oscillating
$\omega_{he}(z)$, we found although the potential $V(\phi)$ shows
an oscillating behavior, this oscillation is different from the
simple sine or cosine function. Here the potential $V(\phi)$ of
the hessence is an oscillating function with the increasing (or
decreasing) amplitude. In last part, we discussed the evolution of
the perturbations of the quintom model, and found the
perturbations of the quintom $\delta_{Q}$ and the metric $\Phi$
are all finite even if at the state of $\omega_{Q}=-1$ and
$\omega_{Q}'\neq0$. We should notice that, in our discussion, we
haven't considered the possible interaction between the quintom
field and the background matter, which may show some new
interesting characters \cite{hao}.

~

\textbf{ACKNOWLEDGMENT}: Zhang Yang's research work has been
supported by the Chinese NSF (10173008), NKBRSF (G19990754), and
by SRFDP. Zhao Wen's work is partially supported by Graduate
Student Research Funding from USTC.

~


\newpage

\begin{figure}
\centerline{\includegraphics[width=15cm]{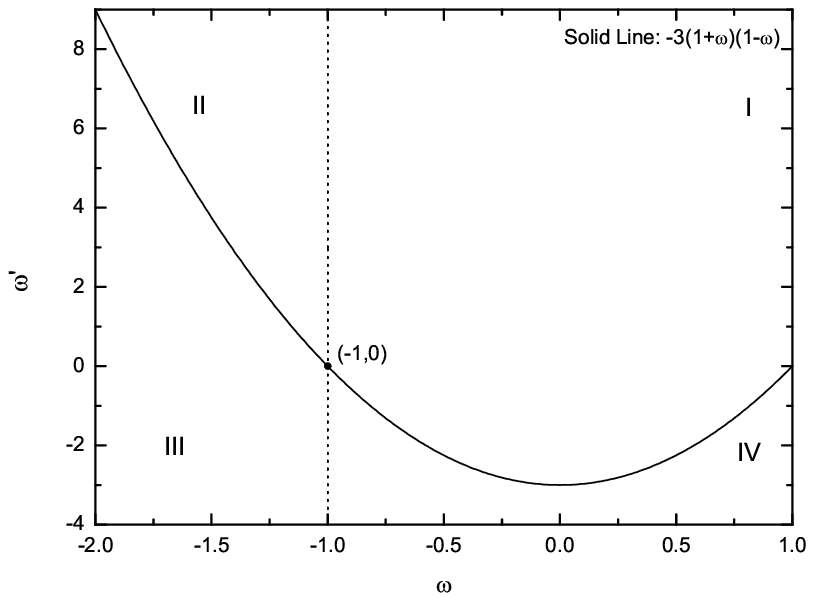}} \caption{\small
The $\omega$-$\omega'$ plane of the hessence model. This plane is
divided into four parts according to the values of $c_a^2-1$ and
$\omega+1$ being larger or smaller than zero.}
\end{figure}

\begin{figure}
\centerline{\includegraphics[width=15cm]{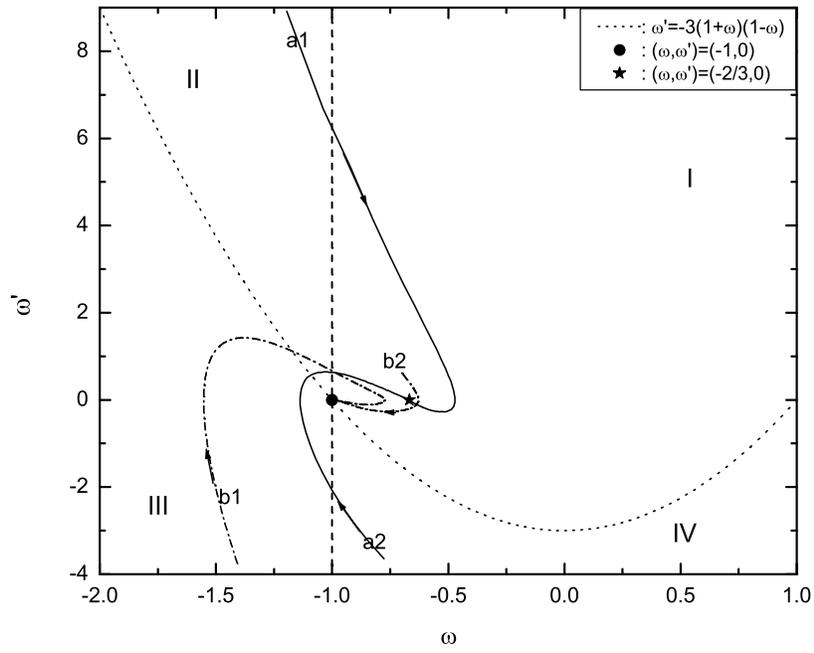}} \caption{\small
The evolution of four hessence models in $\omega$-$\omega'$ plane.
The thin arrows denote the evolutive direction of $\omega$ and
$\omega'$ with time.  }
\end{figure}
\begin{figure}
\centerline{\includegraphics[width=15cm]{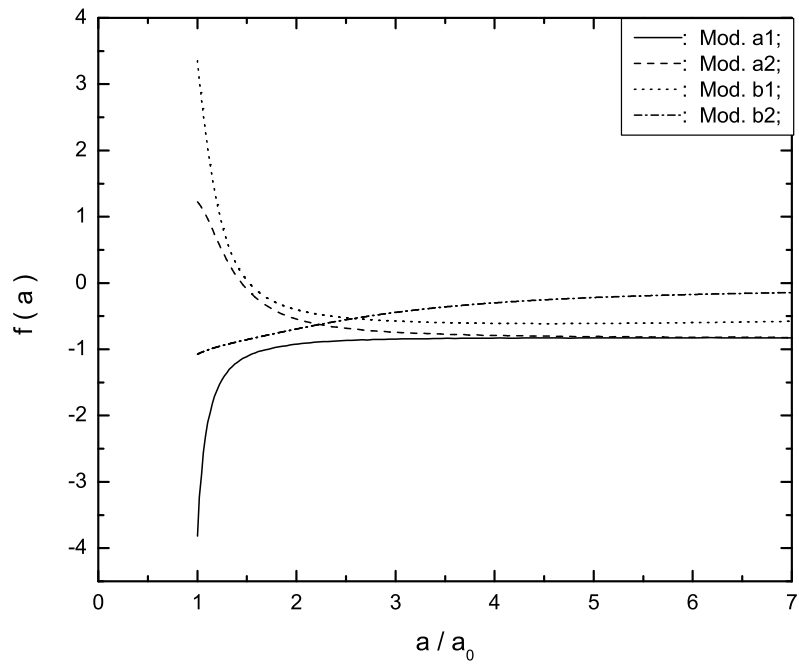}} \caption{\small
The evolution of the function $\dot{V}(\phi)$ in the four hessence
models. Here the dimensionless function $f(a)$ is defined by
$f\equiv\frac{\dot{V}}{H\rho}$. }
\end{figure}

\begin{figure}
\centerline{\includegraphics[width=15cm]{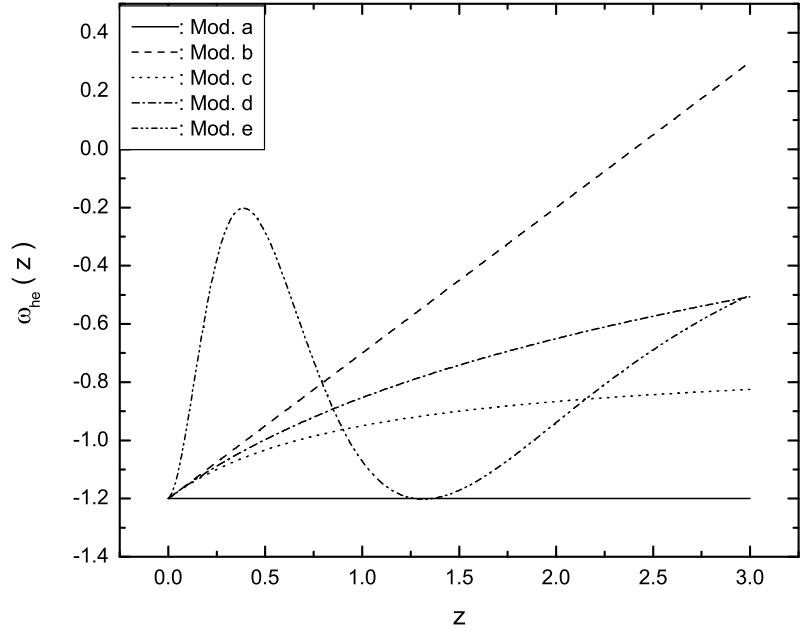}} \caption{\small
The EoS of five kinds of parametrization models.}
\end{figure}

\begin{figure}
\centerline{\includegraphics[width=15cm]{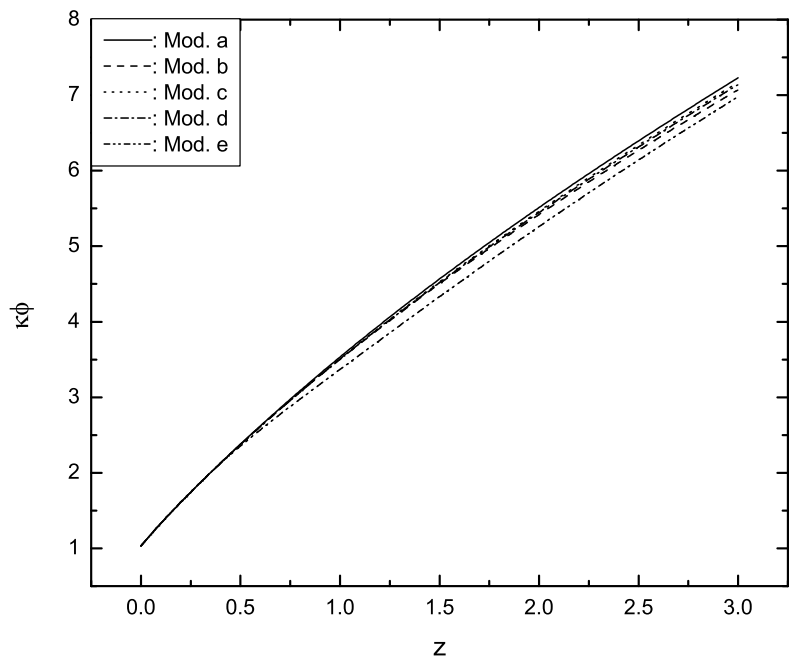}} \caption{\small
Evolution of $\phi$ with the redshift $z$ of the hessence fields.}
\end{figure}

\begin{figure}
\centerline{\includegraphics[width=15cm]{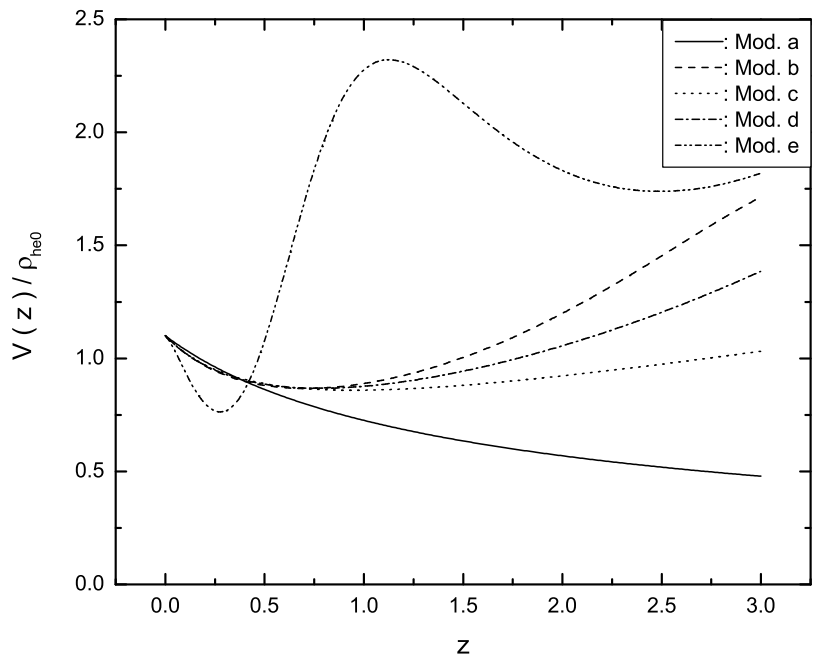}} \caption{\small
Evolution of potentials of the hessence models. }
\end{figure}

\begin{figure}
\centerline{\includegraphics[width=15cm]{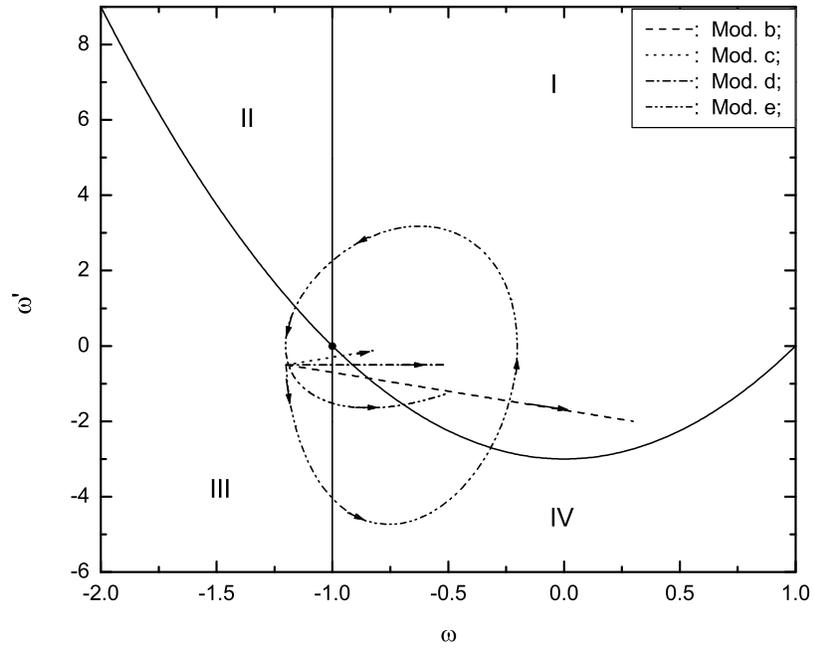}} \caption{\small
The parametrization models in $\omega$-$\omega'$ plane. The thin
arrows denote the evolutive direction of $\omega$ and $\omega'$
with the increasing redshift $z$. }
\end{figure}

\begin{figure}
\centerline{\includegraphics[width=15cm]{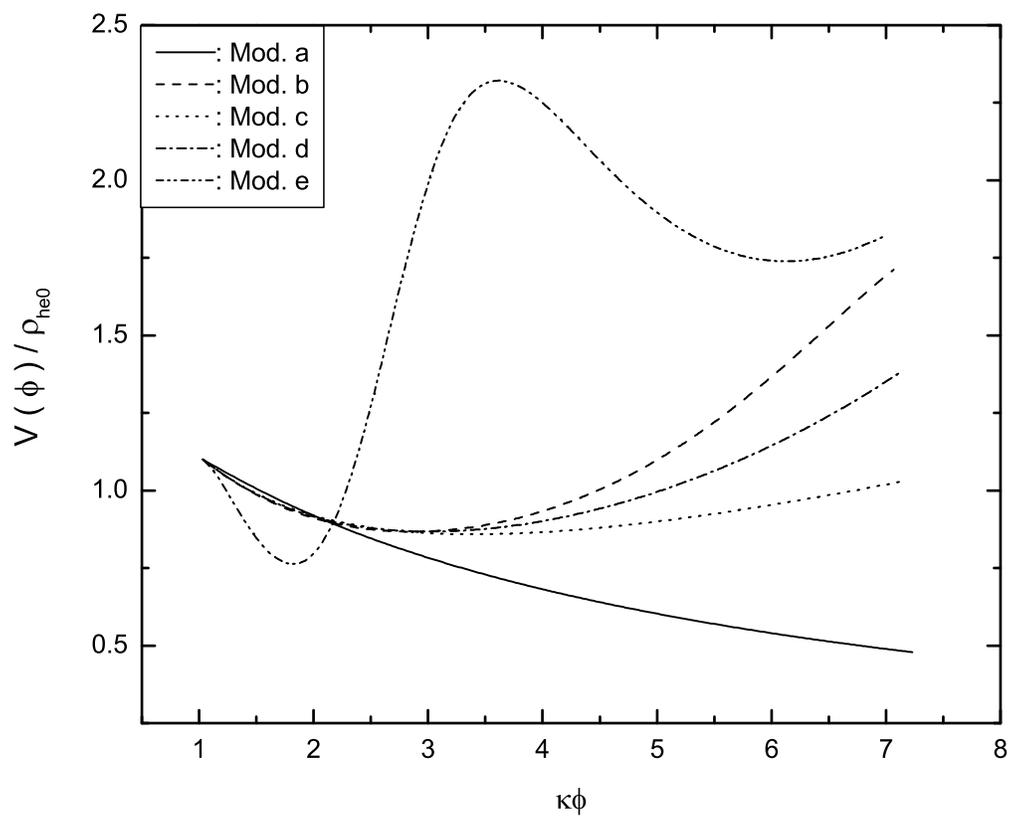}} \caption{\small
Constructed hessence potentials $V(\phi)$. }.
\end{figure}

\end{document}